\documentclass[nonacm,sigconf,screen,9pt,timestamp]{acmart}
\AtBeginDocument{%
  }

\setcopyright{rightsretained}
\copyrightyear{2026}
\acmYear{2026}

\begin{document}

\title{The Essence of Git: Concepts for P2P Replication}

\author{Erick Lavoie}
\correspondingauthor
\email{erick.lavoie@unibas.ch}
\orcid{0000-0002-4020-6578}
\affiliation{%
  \institution{University of Basel}
  \city{Basel}
  \country{Switzerland}
}


\begin{abstract}
Git is most commonly known as a decentralized version control system. Underneath the surface, the design of Git exhibits an elegant object model and incremental replication protocols that have been used to build a variety of applications that go much beyond version control. In this paper, we survey those applications and show how application concepts have been mapped to Git concepts to show the latter's generality. We also show the connection between the Git commit history and the theory of distributed algorithms and convergent replicated data structures, suggesting that it approaches their essence. We finally compare to current peer-to-peer frameworks to establish the level of abstraction of the Git object model and peek at its potential universality. Throughout, we frame our discussion using the \textit{conceptual design} approach proposed by Daniel Jackson.
\end{abstract}




\maketitle

\section{Introduction}

Git was initially designed by Linus Torvald in 2005 as a quick replacement for the BitKeeper distributed version control system, to help quickly merge patches from the large number of concurrent teams working on the Linux kernel subsystems~\cite{brown2018git-origin}. In the two decades since, it has become the \textit{de facto} standard for managing source code. GitHub, a code hosting and collaboration platform that uses Git as a major user interface, reports 180 millions users in 2025~\cite{github2025octoverse} and has become so ubiquitous that it is often conflated with Git itself. What is not as well known, is that Git supports equally well peer-to-peer synchronization directly between repositories. In a sense, a code host such as GitHub mostly acts as a highly available replica and canonical reference.\footnote{It also provides additional collaboration features not natively supported by Git, such as pull-requests, tracking of issues, and access-control but as we will discuss later, there is ongoing work to support similar features without relying on a centralized host (Sec.~\ref{sec:radicle}).}

Over the last twenty years, the Git developers added a large number of additional features, including more efficient storage formats for objects~\cite{git-pack-format} and references~\cite{git-pack-refs-format}, a more efficient synchronization protocol~\cite{git-protocol-v2}, hooks to modify replication behavior with custom scripts~\cite{git-hooks}, etc. All these changes were possible while keeping the original object model from Git essentially stable. Moreover, in parallel, a number of projects have (ab)used the model to implement additional functionalities (ex:~\cite{git-notes,git-meta,git-annex}) and applications that extend and go beyond source version control (ex:~\cite{gloor2025git-monopoly-auctions,arndt2019dkm-on-git,siegfried2025gachix}). Coincidentally, theoretical works around strong eventual consistency have used base concepts that are essentially abstractions of the Git commit history (ex:~\cite{kleppmann2020bec-p2p-db-limits,jacob2024hashchronicle,almeida2025blocklace}). 

De Rosso and Jackson previously published an analysis~\cite{jackson2015conceptual-design} of Git as a version control system, and identified the purposes and misfits behinds its main concepts for the end users~\cite{derosso2013git-conceptual-analysis,derosso2016git-critique}. Jackson later generalized the approach as a theory of \textit{conceptual design}~\cite{jackson2015conceptual-design} and adapted the presentation for a general audience~\cite{jackson2021essence}. In this paper, we reuse the approach~\cite{jackson2015conceptual-design}, namely, we provide a \textit{definition} for key concepts at an abstract level to favor reuse, we identify \textit{the purpose} of each concept to justify its presence, and we provide \textit{operational principles} to illustrate the purpose with user interactions. We however have a different focus: we identify core concepts behind the Git design that enable distributed operation and efficient peer-to-peer synchronization that extend and go beyond source version control, \textit{from the perspective of a peer-to-peer application developer}. Our contributions are to provide a large number of examples to inspire others to use Git in a similar manner, as well as alternative definitions and purposes for the main Git concepts that help thinking of the Git core as a general purpose peer-to-peer framework.

In the rest of this paper, we first present the Git object model, as was originally intended for implementing decentralized version control (Sec.~\ref{sec:background}). We then explain how various applications and designs use it (Sec.~\ref{sec:survey}).  We explain its replication behavior and its relationship to theoretical models (Sec.~\ref{sec:git-theory}). Given this context, we provide an alternative description of the concepts behind the Git object model that better accounts for the full range of applications that use it (Sec~\ref{sec:essence}). We show that the Git object model and current replication protocols are not quite universal but may approach such a status by comparing with ongoing decentralized projects (Sec.~\ref{sec:comp-other-p2p-models}). We finally end with a summary and open questions (Sec.~\ref{sec:conclusion}).

\section{Background}
\label{sec:background}

While Git provides a large number of features, the core object model that underlies these functionalities is surprisingly simple and elegant. It has remained mostly stable since its introduction.\footnote{A "blob", a "tree", a "commit" (named "changeset") and the staging index (named "current directory cache") were there from the very beginning~\cite{torvald2005git}. A 'reference', as well as synchronization protocols and merging behavior were introduced later.} We anchor its presentation with the original purpose of version control.

\subsection{Object Model}

\subsubsection{Preliminaries} A \textit{directory} is a recursive hierarchy of directories (folders) and files on a file system, forming a (mathematical) tree such that directories are internal nodes and files leaves. A \textit{working directory} is a directory that contains the source code of a project that is currently being modified. A development \textit{branch} is the history of revisions to a working directory. A \textit{path} is a sequence of names that identifies either a directory or a file within a directory. \textit{Immutable} objects are objects that cannot be modified after creation. \textit{Mutable} objects are objects that can. 

\subsubsection{Immutable Git objects:} A \textit{blob} (binary large objects) stores only the content of files.  A (Git) \textit{tree} stores meta-data about a list of files (represented as blobs) and directories (represented as trees), including permissions, executable status and paths. A \textit{commit} describes and provides context for specific revisions of the working directory by 1) associating meta-data (author, message, timestamps, etc.) to a snapshot (represented as a tree); 2) causally ordering this snapshot with regard to previous commits by listing the immediately preceding as parent(s). A \textit{tag} provides a meaningful name (ex: release numbers) to specific revisions by associating a path to a commit.

\subsubsection{Mutable Git objects:} A \textit{reference} tracks the history of a development branch by associating a path to a commit. The \textit{staging index} is a hidden directory that contains a copy of the state of a working directory that is intended to be committed.

All objects are tracked in a \textit{repository}, i.e. a special \texttt{.git} hidden folder within a working directory, with the immutable objects (blobs, trees, commits, and tags) in a content-addressed store (\texttt{.git/objects}), and the references in a reference store (\texttt{.git/refs}). Only immutable objects and references are replicated between repositories. Later versions of Git have added support for more objects, (ex: \textit{reflog}~\cite{git-reflog}, a log of all modifications that have been applied to a reference) and correspondingly also track them in a repository (\texttt{.git/logs/refs}). We limit our discussions to those that are used by the projects we survey later.

\subsection{Operations}

There are hundreds of commands that modify the state of a repository. We focus on those relevant for local updates and peer-to-peer sychronization.

\subsubsection{Preliminaries}

A \textit{local} repository is the repository associated to a working directory, on a local machine. A \textit{local} command is one that only modifies the state of a local repository. A \textit{remote} repository is a replica of a local repository that may be modified concurrently. A \textit{synchronization} command is a command that may update the state of a local and/or a remote repository, based on the state of both.

\subsubsection{Local commands} \textit{Committing} is the act of taking a snapshot of the staging index, and 
creating a new commit that references the commit at the tip of the active branch, and updating the associated reference to point to the new commit. \textit{Checkout} is the act of modifying the working directory to reflect the state of a commit and setting it as the new active branch. \textit{Resetting} is the act of modifying a branch's reference to point to an earlier commit in the history, accessible by transitively traversing parents of a commit, and optionally updating the working directory accordingly.

\subsubsection{Synchronization commands} \textit{Pushing} is the act of updating references on a remote repository to reflect the state of local references. \textit{Pulling} is the symmetric equivalent, i.e. updating the state of local references to reflect the state of references on a remote repository. Concurrent modifications of references between a local and remote repository may require \textit{merging}, i.e. creating a new commit that resolves potential conflicts between the two states, or \textit{rebasing}, i.e. computing differences between commits and reapplying them as new commits at the tip of the modified branch. \textit{Pushing} and \textit{pulling} are implemented as connected synchronous protocols, i.e. a repository contacts another repository using an agreed protocol~\cite{git-protocol-v2}, computes and transfers the minimal set of objects that is required for the update, and updates reference(s) of the target repository if necessary. \textit{Bundling} is the act of creating an archive containing the state of references and a subset of commits (and associated objects), potentially up to the full history. A recipient may later asynchronously update their local repository by pulling from the bundle. The bundle may be transferred, e.g., by email or on a USB disk.

\subsection{Accessibility}

The Git conceptual model is simple enough that we\footnote{PhD candidate, Ali Ajorian, and the author.} have successfully taught it for multiple years to 2nd year Bachelor students in Computer Science in the Department of Mathematics and Informatics at the University of Basel. Following the lecture, the students are able to implement a basic chat application in Bash or Python, in a mostly self-driven exercise, with occasional guidance from teaching assistants that have done it the previous year. Some students also quickly learned the object model and relevant operations with no more than partial guidance during a weekly meeting for their Bachelor or Master project (covered in Sec.~\ref{sec:published-designs}).

\section{Survey}
\label{sec:survey}

We now survey works that implement a large variety of applications by adding new objects in repositories as well as by repurposing blobs, trees, commits, tags, and references.

\subsection{Publications}
\label{sec:published-designs}

In this section, we cover published reports, theses, and papers that describe the design of Git-based systems. 

\subsubsection{GOC-Ledger and variants}
GOC-Ledger~\cite{lavoie2023gocledger} is a state-based conflict-free replicated data type (CRDT)~\cite{shapiro2011crdt}, that implements an eventually-consistent ledger using grow-only counters that separately track the total amount of credits ever sent between every directed pair of participants.  The original paper did not discuss any implementation techniques. The Delta-GOC-Ledger~\cite{heisch2026delta-goc-ledger} optimizes the message size by exchanging only the delta-state of accounts, i.e. the minimum set of counters that were modified following a set of operations, rather than the full state of an account on every update. A prototype was implemented on Git~\cite{heisch2026delta-goc-ledger} that tracked each counter value as a blob, and each directed pair of participants as a path in a tree. This approach reused the Git tree merging commands for merging account states. The state of an account could be materialized on the file system with a simple checkout. However, the approach had significant storage overhead: each counter update required two different tree objects, which dominate the overall storage. DOLE v2~\cite{kipfer2026dolev2} adapted the design to encode the delta-state updates directly in a commit meta-data: the sender is the signer of the commit, the receiver is stored in the committer name field, and the updated grow-only counter value is stored in the author email. Compared to the Delta-GOC-Ledger approach, the average size of a transaction when transferred as an update (ex: in a bundle file), decreased from 500B to 110B. In DOLE v2, Git was used strictly as a storage backend but this format still allows two Git repositories to synchronize by quickly exchanging a minimal set of commits using the existing Git replication protocols.

\subsubsection{2P-BFT-Log}

2P-BFT-Log~\cite{lavoie2023bftlog} is a state-based CRDT that implements an append-only log that converges to the longest strictly sequential prefix known across correct replicas. Convergence is guaranteed even in the presence of equivocation, i.e. different updates at the same index of a log that were accepted by different correct replicas in the past. The paper suggests implementing messages as commits, and representing the state of the log as a self-certifying reference, i.e. a reference that includes the public key of the identity used to sign (commit) updates in order to only allow updates from the legitimate author. It did not provide a reference implementation. von Fellenberg~\cite{vonFellenberg2026verification-of-invariants} provides one implementation example and also explains a verification approach that validates both the expected invariants on the 2P-BFT-Log and an embedded Delta-GOC-Ledger payload.

\subsubsection{Package Cache}

Gachix~\cite{siegfried2025gachix} is a cache for Nix~\cite{dolstra2006nix} packages containing binary libraries and executables, implemented on Git. The content of binary executables and libraries is stored as blobs, directories within a package are stored as trees, a package is stored as a commit with its dependencies to other packages stored as parents. The Nix store hash for a package is stored in a reference that points to the package (commit). In addition, Gachix provides transparent translation between current cache protocols used by other tools in the Nix ecosystem and Git operations. Compared to other binary cache implementations, Gachix is more storage efficient, partially due to deduplication offered by the Git object model, and competitive on retrieval latency for small to medium-size packages.

\subsubsection{Games}

Matter implemented the Catan board game~\cite{matter2025git-catan} over Git by mapping the state of every game element to a blob, naming the different parts of the state as paths in a tree, storing the snapshot of the game state after a player action as a commit, and tracking the history of each player's actions with one branch (reference) per player. Each player's sequence of action is an append-only log and causal dependencies between player actions are encoded as extra parents. The game rules sequentially order all player actions, generating a mostly linear history, except for when a seven is rolled: some players may have to discard some cards, which happens in parallel for all players, after which the player who rolled the seven creates a merge commit that merges the resulting state of all the hands of all players. Similarly, Gloor implemented Monopoly~\cite{gloor2025git-monopoly-auctions} but encoded the game state in a YAML file instead of through Git trees and blobs. This greatly simplified (de-)serialization of the state (to and) from Python. While most of the rules of Monopoly sequentially order player actions as well, the occasional auction may trigger several rounds of concurrent bids from players, until a winner is selected among those that did not pass and are still solvable. In effect, the mechanics of the auction implement a consensus algorithm to select the winner. The auction was implemented on top of the append-only logs and provides an example of using Git to implement more complex synchronization protocols than typically used for, e.g. CRDTs.

\subsubsection{Knowledge Management}

The Quit Store~\cite{arndt2019dkm-on-git} is a decentralized knowledge management system, formalized as entity-relationships encoded as \textit{subject-predicate-object} triples following the RDF data format. The Quit Store allows independent development by different teams, while sharing some common base and allowing later merging as consensus emerges on how to model the domain. It externally exposes a standard SPARQL interface, which is a database query language specialized to work with entity-relationships, with additional collaboration operations that are implemented with Git, such as push, pull, or merge. Git commits, and associated trees and blobs, are used to record both the changes to the knowledge graph, in the form of additions and removals of RDF triples and the SPARQL query that generated them, and the resulting graph after applying the changes. Each new version of a graph is stored in a blob and benefits from the delta encoding offered by the Git optimized pack format~\cite{git-pack-format} to lower storage requirements.

\subsubsection{Text Editor}

Pinnacle~\cite{Chernyakhovsky2012Pinnacle} is a design concept for a collaborative paragraph-oriented text editor that uses Git to track version history, and operational transforms to resolve concurrent modifications to the same paragraph. The content of paragraphs is stored in blobs, unique identifiers for paragraphs are stored as paths in trees that point to the relevant blobs, and a reserved table-of-content tree path points to a blob containing the sequence of paragraph identifiers. Local text operations performed by each user are stored outside of Git and resolved through a leader-based external protocol or word-by-word three-way merge strategy. 

\subsubsection{Peer Review}

Git-reviewed~\cite{mukadam2014git-peer-review} is a decentralized tool for peer-reviewing code tracked with commits. The reviews are also stored in Git and each commit may be the target of multiple reviews. The set of all replicated reviews is tracked under a single branch, and each commit of that branch represents the set of reviews tracked at that point in time. At any given point, all reviews for all commits are stored in a single global tree, in which the targeted commit identifier is used as a path, and the set of associated reviews stored in a sub-tree. The sub-tree stores each individual (immutable) review as a blob using the blob identifier as a path as well, effectively using a tree as a set. This guarantees that reviews created concurrently will always be successfully merged in the same set with Git's merge command. The design was modeled after the mailing list discussions used to review commits on leading open source projects, such as the Linux Kernel, in which emails are also immutable after being sent. 

\subsubsection{Access Control}

Walser~\cite{walser2025ssh-ac} presents an access-control mechanism for remote Git repositories accessible through SSH~\cite{ietf2006ssh-architecture} where permissions are given and revoked based on transitive trust relationships between users, using a trust graph similar to Secure-Scuttlebutt~\cite{kermarrec2021ssb-gossiping}. The transitive trust relationship has a static relationship to the permissions: the owner of a repository always has access to the repository, the users directly trusted by the owner have read and write access, the users indirectly trusted by the owner by at most one level have read access, and all other users have no access. The operation of adding or removing trust is encoded in the meta-data fields of a commit, and the commit is authenticated by the user. All commits from the same user form an append-only log. Because access only depends on the presence of a path in a trust graph between the owner and the target user, and any user is always allowed to publish/revoke his trust no matter whether they are trusted or not, no causal links between user updates are stored. When trust signals are modified and valid, the trust graph is updated within the owner's repository and a Git hook script automatically updates the access permissions in the SSH authorized keys on the owner machine.

\subsubsection{Git for Large Datasets}

XetHub~\cite{low2023xethub}  transparently maps large files to a content-defined merkle-tree decomposition which provides automatic deduplication when small parts are modified. The internal nodes of the merkle-tree are represented as Git trees. Chunks of the files are stored as external blobs in an external content-addressed store.  External references to external blobs are stored in (Git) blobs. Modification to the set of all merkle trees for an entire repository is incrementally tracked with Git notes (Sec.~\ref{sec:git-notes}) enabling deduplication across branches. The conversion from the working directory to the internal XetHub format during commit and checkout is implemented as pre-commit and post-checkout scripts. Commits and references are still used to represent revisions to datasets.

\subsubsection{Secure Git}

Li \textit{et al.}~\cite{li2025git-end-to-end-encryption} show how to secure Git repositories by: 1) making the content of files and later updates private (\textit{confidentiality}); 2) preventing adversaries, including the hosting server, from updating the repository (\textit{repository unforgeability}); and 3) verifying that each version has not been modified (\textit{integrity}). The Git object model is still used for its original purpose with the difference that the content of files in blobs is encrypted, either 1) on a per-line basis, or 2) on a per-character basis with all changes in a given file collected together. Their scheme is compatible with existing code forges, such as GitHub and GitLab, since encryption and decryption is performed within local repositories. Compared to previous approaches also encrypting revisions, they achieve both smaller update sizes and storage requirements, by encrypting only the modifications in newer revisions  which makes the encryption overhead proportional to the update size. The tradeoff is that there is no confidentiality on which files were modified in each commit.

\subsubsection{Gitless}
\label{sec:gitless}
 Gitless~\cite{derosso2016git-critique} is a different command-line front-end for Git, that simplifies the development workflow by removing the need for a staging index, simplifying the management of tracked (version-controlled) files, storing merge conflicts as first-class objects (commits), and automatically saving and restoring the state of uncommitted files when switching branches. All these changes are still compatible with the normal behavior of blobs, trees, commits, tags, and references, therefore the resulting repository is interoperable with Git repositories and code forges. Gitless predates the apparition of Jujutsu~\cite{jujutsu}  (Sec.~\ref{sec:jujutsu}) by 6 years and the design is the first published example of applying the conceptual design approach of Jackson to a concrete project~\cite{jackson2021essence}.

\subsection{Open Source Projects}

The following projects are actively developed but have no associated publication on their design. The following summaries have been assembled from the available documentation and experimenting with the tools.

\subsubsection{Jujutsu}
\label{sec:jujutsu}

Jujutsu~\cite{jujutsu} is a distributed version control system, which was started in 2019. It is backward-compatible with Git, because it uses the same store formats and synchronization protocols, but tailored to a different workflow.  Among the notable differences~\cite{jujutsu-git-comparison} are: 1) the automatic commit of the current state of the working directory, removing the need for a staging index; 2) a workflow organized around \textit{changes}, i.e. accumulated differences compared to a previous commit, without the need for naming associated references; 3) first-class tracking of conflicts; 4) tracking of the evolution of the entire state of the repository -- including references, unreferenced commits, checked-out branches (heads), state of working directories -- with an operation log~\cite{jujutsu-concurrency-operation-log}, enabling undoes on any operation.  

Non-standard behaviors are implemented as follows: the state of the working directory is saved as a normal commit with a stable \textit{change identifier} stored as extra meta-data in the commit message that carries across new versions; conflicts are stored as commits, similar to Gitless~(Sec.~\ref{sec:gitless}); the operation log is maintained locally in an operation store on the file system, but is not shared between jujutsu repositories. An operation store could however also be shared through Git, as done by GitButler~\cite{git-butler-operations-history}, a similar concurrent project.

\subsubsection{Git notes}
\label{sec:git-notes}

Git-notes~\cite{git-notes} are used to comment on an existing immutable object without modifying it, most often targeting commits. The history of notes modification is stored under a single \texttt{notes/commits} reference for the entire repository. The state of all notes is stored in a tree within a commit. Each tree stores a target object identifier as a path and the note content for it as a blob. Because a tree may have only one object associated to a path, there can be at most one note targeting each object. Notes can be replicated as any other branches with the usual synchronization commands.

\subsubsection{Git-meta}

Git-meta~\cite{git-meta} is a specification, and reference implementation, to exchange meta-data about Git objects and repository concepts, such as the provenance of data, ownership, attestations, etc. It supersedes Git notes with a richer meta-data model. The Git object model is used to encode meta-data tuples $(\textit{target}, \textit{key}, \textit{value})$ where the target is the Git object or repository concept, the key is a user-defined arbitrary string, and the value is either a string, an (unordered) set of strings, or an (ordered) append-only list of strings. The meta-data is stored locally in an sqlite database but is serialized as Git objects for replication. Similar to Git notes (Sec.~\ref{sec:git-notes}), all meta-data tuples are serialized in a single tree, committed and tracked under a reserved reference. A path within the meta-data tree encodes both the target, the key, and the value type of the meta-data tuple. The associated object depends on the value type: a string is a blob; a set of strings is a tree in which both the path and associated blob use the same blob identifier; a list of strings is a tree in which the path is a combination of the timestamp (time at which the entry was appended) and identifier prefix of the associated string (blob). All value types are state-based CRDTs~\cite{shapiro2011crdt} with automatic merge semantics.

\subsubsection{Radicle}
\label{sec:radicle}

Radicle~\cite{radicle-protocol-guide} is a peer-to-peer code collaboration network, started in 2019~\cite{radicle2026history}, with features such as tracking and hosting Git repositories, as well as per repository tracking of issues, pull-requests (called \textit{patches}), and maintainers. The first iteration was based on IPFS~\cite{benet2014ipfs} but was too slow to use for regular development and lacked efficient support for mutability of replicated objects~\cite{radicle2026faq}. Since the third major iteration in 2022, the system was reimplemented to use Git itself and is sufficiently mature to host its own development. 

Each repository hosted on radicle is stored in a meta-repository that behaves like a normal repository but reserves certain references to track additional concepts. Within a meta-repository, each hosted repository is stored under a different namespace~\cite{git-namespaces}, i.e. reference prefix, but all namespaces share the same immutable object store, enabling deduplication across repositories. The network identity for a repository is obtained from the identifier of the blob that stores an identity document that contains the identifier(s) of the initial owner(s), a threshold of owners required to make a modification, a name and description for the repository, and the name of the default (main) branch.  Ownership is updated by tracking newer versions of the identity document in a commit chain, forming an append-only log rooted in the original identity document, in which an update is only valid if it was signed by at least a threshold of valid owners from the previous update. The latest update of the chain is referenced under \texttt{refs/rad/id}. Operations on issues, patches, and maintainers are tracked using a Git commit history that encodes the causal dependencies between concurrent operations and the latest state of each is referenced under \texttt{refs/cobs/xyz.radicle.<object>/<id>}. To make updates to a repository, a developer clones the repository using the rad tools, which will retrieve usual repository objects but leave out meta-objects (i.e., repository identity, ownership modification chain, issues, patches, and maintainers), performs normal Git operations on the clone, then push the changes back to the meta-repository with the radicle tools. To update any of the radicle-specific concepts, the radicle tools is used which will transparently update the meta-repository.

\subsubsection{Git-annex}

Git-annex~\cite{git-annex,hess2025gitannex} is a decentralized file manager, started in 2010~\cite{hess2026thanks}, that transparently tracks a subset of files as "annexes" to an existing Git repository, essentially an external blob management system. Regular development branches may store the identifiers of annexed files that are transparently linked to content stored inside the annex object store when checked out.  A collection of over one hundred commands helps manage the annex, including the distribution of the actual content through almost every protocol and storage solution currently available, peer-to-peer or cloud-based. Git-annex is typically used to manage collections of large files but works equally well to coordinate redundant backups for directories over a variety of medias and platforms. 

Each repository is assigned a unique identifier number and tracks files with a CRDT, stored in the \textit{git-annex} branch in a commit history, that logs the date, location (i.e. repository identifier), and availability of external blobs on all participating repositories. Using the data structure, replication policies can be automatically enforced, such as keeping a minimum of 2 but no more than 3 copies across all tracked repositories. 

Advanced use cases using \textit{special remotes} include: 1) computing derived content on-demand from existing replicated content using remote computation resources, e.g. to provide media content in a variety of formats, build packages from sources, etc., (a step towards asynchronous \textit{named function networking}~\cite{tschudin2014named-function-networking}); 2) interoperability with git-lfs (Git Large-File-System) enabling the use of GitHub and Gitlab as storage providers, Amazon Glacier for cheap cloud archival, or the Internet Archive for long-term file archival; 3) defining new special remotes following a key-value protocol~\cite{hess2013external-special-remote-protocol} based on adding, removing, and checking the presence of a file under a given key.

\subsection{Discussion}

The Git object model was originally designed as a model of a versioned file system: blobs, trees, commits capture a specific snapshot with meta-data, while tags and references serve as a directory of versions. A file system is itself a model that has been widely adopted across operating systems to serve as an abstraction mediating between end users and application developers on one side, and kernel and hardware developers on the other. Moreover, because the implementation of Git is itself using the file system to maintain the state of a repository with stable formats and well-defined incremental operations, it has been possible to extend it and build new applications in a variety of programming languages. So the breath of extensions and applications for Git can be partially explained by the fact that it models and is implemented using the concepts of a file system that are well established and independent of programming environments.

%

\section{Replication Behavior and Theoretical Underpinnings}
\label{sec:git-theory}

The Git object store behaves as a replicated set of immutable objects. If objects are never deleted, it behaves as a grow-only set CRDT~\cite{shapiro2011crdt}. The \texttt{git fetch} command, when given the identifier of a remote immutable object, ensures one is replicated in the local store. This works whether the identifier of the object is provided directly\footnote{There is no symmetric "push" equivalent: the \texttt{git push} command may only specify references to update remotely, not immutable objects to replicate. This can be circumvented by lifting the object identifiers to the tag level, e.g. adding a \texttt{refs/tags/objects/<id>} for each object \texttt{<id>} and validating consistency during a fetch or push.} or transitively because it is referenced inside another object or by a Git reference.

The Git reference store has richer and more complex replication behavior, with complex merging behavior that may automatically create new commits when references have been concurrently modified on two repositories. This merging behavior can be leveraged or customized to implement more complex CRDTs, such as causal-length set~\cite{weihai2020causal-length-set} or byzantine fault-tolerant append-only logs~\cite{lavoie2023bftlog}. 

All modern distributed abstractions~\cite{cachin2011intro-distr-prog}, including consensus, replicated registers, and view changes, can be implemented on top of Git using the following mapping: a message is implemented as a commit; point-to-point delivery corresponds to copying a commit from one repository's object store to another one with the application being notified only if the commit was not there already; and the fact that a message \textit{happened before}~\cite{lamport2019causal-time} another is attested by the former being a parent (or ancestor) of the latter. The assumption that participants behave sequentially requires enforcing that their commit history behaves as an append-only logs~\cite{lavoie2023bftlog}. Using this mapping, a set of (authenticated) repositories may  automate management tasks, e.g., to coordinate when some objects or references may be safely purged from all stores, and a set of (authenticated) users may coordinate their behaviour, e.g. use consensus to perform an auction~\cite{gloor2025git-monopoly-auctions}.

More recent theoretical work on strong eventual consistency~\cite{shapiro2011crdt} in the presence of Byzantine faults uses an abstract version of a commit history, which retains the signature for authentication and parent links for causality and abstracts all other fields as application-specific \textit{payload}. This is known under various names such as hash chain~\cite{kleppmann2020bec-p2p-db-limits}, blocklace~\cite{almeida2025blocklace}, or hash chronicle~\cite{jacob2024hashchronicle}. Enforcing sequentiality of commits from the same author and converging to the longest sequential prefix in the presence of equivocation, results in a Byzantine fault-tolerant append-only log~\cite{lavoie2023bftlog}, and delaying the replication of commits until equivocation is acknowledged results in Byzantine repellance of equivocators and colluders~\cite{almeida2025blocklace}. 

There are still open questions about how to build applications within a byzantine eventually-consistent model~\cite{kleppmann2020bec-p2p-db-limits} because it falls outside the two main models that are known for building distributed applications. On the one hand, in the usual CRDT model~\cite{shapiro2011crdt}, in which the state of an application monotonically grows over an imposed semi-lattice, an update is either included and forever after reflected in the state of an application or outright rejected. On the other hand, in typical distributed byzantine fault-tolerant abstractions~\cite{cachin2011intro-distr-prog}, much more expensive synchronization primitives are instead used to hide intermediary states from the application when there is uncertainty about the decision, to ensure the decision is final and coherent on all replicas once made. However, when designing for strong eventual-consistency at the meta-level, i.e. on systems that enable or disable some operations for users based on the current replicated state, some decisions may be reverted after more messages are replicated, e.g. when a user is/isn't permitted to change the permission for another user~\cite{jacob2026systemorientedformalverificationlocalfirst} or when a user had/hadn't enough credit in their account to perform an operation~\cite{lavoie2024closed-knowledge-commons} because the funds turn out to have been double-spent~\cite{lavoie2023gocledger}. Once this is better understood, the sufficiency of the Git object model or a search for new necessary concepts could be undertaken.

%

\section{Essence of Git}
\label{sec:essence}

We now revisit the object model and operations previously presented in Section~\ref{sec:background}, in light of the more general purposes exhibited by all the designs we surveyed and the currently understood theoretical underpinnings. For concepts that can benefit, we provide a more general purpose that abstract version control and example interactions that capture the intent (i.e. \textit{operational principle}).

A \textit{path} is a name defined by an application to group objects hierarchically, with each level separated by a '/'. Querying a prefix of a path returns the directory of objects sharing the same prefix.

A \textit{tree} stores the state of an application as a directory by associating each sub-state (recursively stored as a tree, blob, commit or tag\footnote{Git allows commits and tags to be the targets of paths in a tree.}) to a different path. When mapping two different states as trees, the sub-trees that are identical reuse the same objects (\textit{sub-tree deduplication}). 

A \textit{blob} stores arbitrary content, in a binary or text format. Storing the same content twice will result in the same blob object. Storing two different contents that differ only by a small amount may result in the second eventually being stored as a delta modification of the first (\textit{content deduplication}). 

A \textit{commit} stores an application update. Each update of an application may generate a new commit; retrieving a previous commit allows reconstructing the state of the application at the time the commit was created. The \textit{payload} of a commit (message, fields, and/or a tree) may store update details such as who, what, when, where, for what, and the result on the application state. The \textit{parents} of a commit are previous updates (commits) that are causally related. Retrieving all commits transitively reachable from parents provides the causal history of an update and may be used to validate and causally order updates. The \textit{signature} of a commit authenticates the updater. Commits authenticated by correct updaters (that do not share their private key), may only have been created by them.

A \textit{reference} associates a path to a commit to identify a shareable current (sub-)state of the application. Synchronizing a reference updates the (sub-)state of an application to the latest of the local and remote states, either locally (when \textit{pulling}) or remotely (when \textit{pushing}). When a reference is concurrently modified locally and remotely, this may require resolving the concurrent modifications to obtain a new valid state (\textit{merging}) and associated commit. \textit{Remote} references are local replicas of the state of references on a remote repository at the time of the latest synchronization.

A \textit{repository} tracks the state of an application. A repository has multiple stores that track objects that influence the state. Objects in a \textit{private store} (ex: Git configuration and application-specific stores) are never shared with other repositories and only influence the local state. Objects in \textit{public stores} (ex: immutable objects and references) may eventually influence the state of remote applications.

Other concepts retain their usual meaning. The power of this application model is that it abstracts most of the low-level details of storage and synchronization protocols, reuses most of the work that has been put in developing reliable and efficient tools for version control including the supporting infrastructure (GitHub, GitLab, etc.), and provides a familiar file system abstraction. Moreover, the choice of synchronizing with a single (conceptual) repository, as commonly done with code forges such as GitHub and GitLab, or replicating in a fully peer-to-peer manner is entirely up to the final users, the model equally supports both.

\section{Comparison to Other P2P Models}
\label{sec:comp-other-p2p-models}

Many peer-to-peer projects focus on how to replicate messages between peers with more or less sophisticated reliability, filtering, routing, and ordering guarantees but provide no object model to define the state of an application. We present a somewhat arbitrary selection. Secure-Scuttlebutt~\cite{tarr2019ssb} is organized around append-only logs, with one authenticated log per user, all application messages multiplexed within the same log, and replication based on subscriptions between users~\cite{kermarrec2021ssb-gossiping}. Nostr~\cite{nostr} is based on best-effort delivery of messages, with messages replicated from one user to the other, if they both interact through at least one common relay server, the latter user subscribed to the messages of the first, and the relay correctly cooperates. P2Panda~\cite{p2panda} provides a set of libraries that can be used to obtain replicated append-only logs similar to Secure-Scuttlebutt, possibly augmented with additional access-control and privacy guarantees. In comparison, Git's model is at a higher level because it models a versioned file system, which would be considered an application for the previous projects.

In contrast, Willow~\cite{willow2026datamodel} provides a state-based object model closer to Git. Each message is the combination of a payload, path, timestamp, sub-space (e.g. author), and namespace. Newer messages (i.e. with a larger timestamp) appearing at a prefix of a path supersede all earlier messages at a path suffix within the same sub-space and namespace. Compared to Git commits, the causal relationship between messages is not explicitly encoded (but an application may still add it to the payload) and a given message may appear only under a single path. This approach enables easier deletion of history and efficient synchronization through range-based set reconciliation~\cite{meyer2023range-based-set-reconciliation}. The difference with the Git object model and associated synchronization protocols show that the latter is not quite universal but the proximity of the concepts hints at what a universal model may be.

We believe the Willow object model could be emulated on Git with a combination of commits (without parents) and references, and set reconciliation with similar incremental behavior could be implemented using the existing Git synchronization protocols with replicated index trees. Fleshing out and validating such a mapping will be the subject of future work.

\section{Conclusion}
\label{sec:conclusion}

We have presented a large variety of applications that target the Git object model and reuse its replication protocols, which showed their  usefulness much beyond version control. We also showed the connection between the Git commit history and abstractions used to define  widely known distributed algorithms and general convergent replicated data types, suggesting it approaches the essence of both. Following both observations, we proposed alternative definitions for the key concepts that abstract version control and better emphasize the capabilities of Git concepts for implementing distributed applications \textit{that allow the end-user to decide with whom and when to synchronize}, be they intermittently available peers or highly available and trusted hosts.

We conclude with a list of open questions, distilled from previous discussions:
\begin{itemize}
	\item Are object models and replication protocols of ongoing peer-to-peer projects (for projects that define both) equivalent in expressivity and efficiency to Git? 
	\item What object models could be candidates as a foundation for peer-to-peer replication, analogous to the Zermelo-Fraenkel set theory that the mathematics community has adopted as a common foundation of mathematics? Is the Git object model and the associated replication protocols sufficient?
	\item What are the best concepts to base an object model on and associated replication protocols, so that they are convenient for building peer-to-peer byzantine fault-tolerant applications? What are the minimal extensions (if any) that would be required to make Git so?
\end{itemize}

We finally note that designing distributed applications so that they are resilient to malicious peers while leaving the choice of whom to synchronize with to end users, by definition, allows users to recover from compromised centrally-managed hosts. More emphasis on this model as a design target would lead to more resilient computing infrastructure while achieving a better balance between the interests of users and those of platform and service operators.

\bibliographystyle{ACM-Reference-Format}
\bibliography{essence-of-git}

\end{document}